\documentclass[12pt]{elsart}
\usepackage{epsf}
\usepackage{amsmath}
\usepackage{amssymb}

\newcommand{\PSfigure}[2]{%
    \centerline{\epsfxsize=#2\textwidth\epsffile{#1}}
}

\begin{document}
\bibliographystyle{elsart-num}

\begin{frontmatter}

\title{Diffraction theory and focusing of light by left-handed materials}

\author{A. L. Pokrovsky},
\author{A. L. Efros},

\address{
    Department of Physics \\
    University of Utah\\
    Salt Lake City, UT 84112 \\
    USA }

\thanks{%
    This work supported by the NSF grant DMR-0102964.
}


\begin{abstract}
A diffraction theory in a system consisting of left-handed and 
right-handed materials is proposed. The theory is based upon the Huygens's principle and 
the Kirchhoff's integral and it is valid if the wavelength is smaller than any relevant
length of the system. The theory is applied to the calculation of the smearing of the
foci of the Veselago lens due to the finite wavelength. We show  that the Veselago lens
is a unique optical instrument for the 3D imaging, but it is not a ``superlens'' as it 
has been claimed recently.
\end{abstract}

\end{frontmatter}

In his seminal work Veselago \cite{ve,ve1} has introduced the concept
of left-handed materials (LHM's).
In a simplest case the LHM's are materials 
with simultaneously negative
electric permittivity $\epsilon$ and 
magnetic permeability $\mu$ in some frequency range. 
In the LHM the vectors ${\bf k}, {\bf E}, {\bf H}$
form a left-handed set, while in the usual materials ($\epsilon > 0$, $\mu > 0$)
they form a right-handed set.
If imaginary parts of $\epsilon$ and $\mu$ are small,
the electromagnetic waves (EMW's) propagate in the LHM but they have some 
unusual properties. All these properties originate from the fact that
in the isotropic LHM the Poynting vector 
${\bf S} = {\bf E} \times {\bf H}$ is anti-parallel to the 
wave vector ${\bf k}$.

\begin{figure}
    \PSfigure{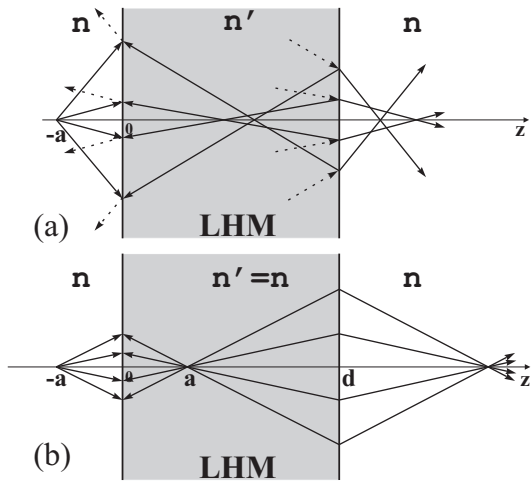}{0.5}
    \caption{%
    Reflection and refraction of light outgoing from a 
point source at $z=-a$ and passing through the slab of the LHM at $0<z<d$.
Refraction of light is described by the anomalous Snell's law.
The arrows represent the direction of the wave vector.
The reflected waves are shown by dashed lines near each interface only.
The slab is surrounded by the usual RHM. (a) $n' > n$. 
(b) The Veselago lens. The reflected waves are absent, 
all rays pass through two foci.
    }
    \label{fig1}
\end{figure}

Consider a  propagation of the EMW from a point source located 
at the point
$z = -a$ through an infinite slab of the 
LHM with the thickness $d$ and a 
usual right-handed material (RHM) at $z<0$ and $z>d$ (Fig.\ref{fig1}).
It is obvious that $S_z > 0$ everywhere at $z > -a$
because the energy propagates from its source.
The directions of vector ${\bf k}$ for different rays are shown by  arrows. 
They  should be  chosen in such a way that at both
 interfaces tangential components of vector ${\bf k}$
for incident, reflected and refracted waves are the same. Another condition is 
that the component $k_z$ should be parallel to $S_z$ in the RHM and anti-parallel
in the LHM.
Then in the LHM $k_z$ is negative.
It follows that the Snell's law for the RHM-LHM interfaces
has an anomalous form:
$\sin i / \sin r = -n'/n$, where $i$ and $r$ are the angles of incidence and
refraction respectively, $n' = \sqrt{|\epsilon '| |\mu '|/\epsilon_0 \mu_0}$ 
and $n = \sqrt{\epsilon \mu/\epsilon_0 \mu_0}$ are positive refractive indexes
for LHM and RHM respectively. 
The angles of reflection are equal to the angles of incidence.

The device shown at Fig.\ref{fig1}(b) is a unique 
optical lens proposed by Veselago. 
In this lens  $\epsilon = -\epsilon '$ and $\mu = -\mu '$, then 
$n' = n$ and $i = -r$.
It is easy to show that in this case the reflected wave is completely absent.
Since all the rays going right from the source have $i=-r$,
all of them have foci at points $z=a$ and $z=2d-a$ 
as shown in Fig.\ref{fig1}(b).
 
All the ideas above have been put forward by Veselago 
about $35$ years ago\cite{ve,ve1}.
Recently the method of fabricating of the LHM's on the
basis of metallic photonic crystals has been found and 
the San Diego group has reported the first observation
of the anomalous transmission \cite{sm2} 
and even the anomalous Snell's law \cite{sm3}.
Both observations have been interpreted as a result
of negative $\epsilon$ and $\mu$.

One can see that the Veselago lens, shown at Fig.\ref{fig1}(b), 
is an {\it absolute instrument} because it images 
stigmatically a three-dimensional domain $-d \le z \le 0$ and the 
optical length of any curve in the object space is equal to the optical length of
its image\cite{born}.
Note, that the definition of the absolute instrument assumes 
geometrical optics only.
Since the LHM's have been already obtained  we think that the 
Veselago lens can be extremely important device  for the  
3D imaging.

Pendry \cite{p2} claims that the Veselago lens has another unique property.
Due to Pendry the resolution of this lens does not have a 
traditional diffraction limitation which follows 
from the uncertainty principle.
Pendry has introduced a new term  ``superlenses'' with the Veselago 
lens as
a first representative of this class. 
Several works have recently appeared where the concept of superlenses 
was criticized \cite{com1,com2,va,nv}. On the other hand, 
Ziolkowski and Heyman \cite{ziol} proposed a refined version of
Pendry's analysis which confirms his result.
Haldane \cite{hal} interpreted results of Pendry in terms
of surface waves described  by Ruppin \cite{rup}. 
Haldane also proposed   a way how to overcome some mathematical inconsistencies
of papers \cite{p2,ziol}, shown before in \cite{lens,nv}. 
Haldane claims that even though the perfect focusing is impossible
the diffraction limit can be defeated.

In this paper we propose a general scalar theory of diffraction in the LHM which
is based upon the Huygens's principle and the Kirchhoff's integral. 
As any diffraction approach our
theory works under  condition that the wavelength is much smaller
than any relevant geometrical length in the problem.
We apply this theory to the Veselago lens and calculate the smearing 
of the foci due to the finite wavelength.
Thus, our result does not support the idea of  ``superlens''.

The Green function for the Helmholtz 
equation for the LHM, which describes propagation of a 
spherical wave from a point source, has a form
$\exp{(-i k R)}/R$, where $k=\omega n/c$ and  $R$ is a distance from the source.
At a small element of the sphere $R = const$ the spherical wave 
can be considered as a plane wave which is characterized by 
the Poynting vector ${\bf S}$ and wave vector ${\bf k}$ both with  the radial
component only.
Since ${\bf S}$ is directed along the external normal to the 
surface element, the wave vector ${\bf k}$ in the LHM is directed 
along the  internal normal.
It is easy to see that our Green function obeys these properties.

Following the principles of the scalar theory of diffraction\cite{lan}
the field $u$ at the observation point $P$ can be written in a form
of a surface integral
\begin{equation}
\label{a}
u_P =  b_l \int u \frac{e^{-i k R}}{R} d f_n,
\end{equation} 
where $R$ is the length of the vector from the point $P$ to the surface element,
 $d f_n$ is the projection of the surface  element $d f$
on the plane perpendicular to the direction of the ray coming 
from the source
to $d f$, $b_l$ is a constant for any LHM.
To find  $b_l$ one can consider a plane wave with the
wave vector normal to the infinite plane of integration. 
Since this plane is 
fictional, the constant can be found from the condition that 
the Huygens's principle in the form Eq.(\ref{a}) 
reproduces the same plane wave.
Doing the  calculations similar to  Ref.\cite{lan} 
one gets $b_l = -k/ 2 \pi i$, so that
for the LHM the constant  $b_l$ has a different sign than 
the similar constant $b_r$ for the RHM.

The Huygens's principle can be applied to any interface which
has a curvature larger than the wavelength.
It gives the correct direction of refracted waves but it does not 
give the amplitudes of both refracted and reflected waves.
However, it can be successfully applied to the Veselago lens 
where reflected waves are absent.

Note, that there are other methods to describe 
the diffraction which may be also used 
if the source of the rays is unknown.
They are described and compared in details in 
Jackson's textbook \cite{jac}.
One can show that all the methods give the same result at $r=-i$.

Moreover,  the Kirchhoff integral Eq. (\ref{a}) may be considered
as a hint to get a solution, valid at large $k$. 
The function $u$ as given by Eq. (\ref{a}) is meaningful only at 
the distances from the interface larger than the wavelength. Nevertheless, 
if we consider solution obtained for these distances exactly at the interface,
we find that it satisfies proper boundary conditions. 
Namely, $u$ is continuous and $\partial u/\partial z$ 
changes sign at the interface (See\cite{ziol}).

\begin{figure}
    \PSfigure{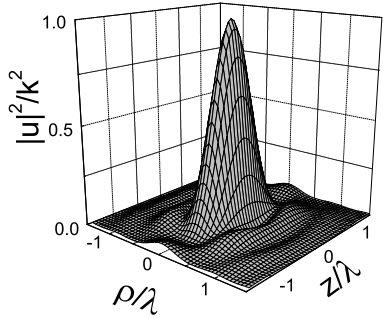}{0.5}
    \caption{%
Distribution of the dimensionless square modulus of the scalar field $|u|^2/k^2$ near the foci of the Veselago lens as a function of $\rho$ and $z$ as
given by Eq.(\protect\ref{c}). Here  $\lambda=2 \pi/k$ is the wavelength.
    }
    \label{fig2}
\end{figure}

Now we apply Eq.(\ref{a}) to the Veselago lens.
To find the field $u$ inside the slab we shall 
integrate in Eq.(\ref{a})
over the plane $z=0$. The field $u$ in this plane is
produced by a point source and has a form 
\begin{equation}
u(x, y, 0) = \frac{e^{i k \sqrt{a^2 + x^2 + y^2}}}{\sqrt{a^2 + x^2 + y^2}}.
\end{equation}
The field inside the slab can be found using Eq.(\ref{a}) with a 
constant $b_{rl}$ instead of $b_l$ because now we are 
integrating over the 
 RHM-LHM interface rather than over the fictional surface in the LHM.
In a similar way at the LHM-RHM interface one should use a constant $b_{lr}$.
Using the method described in Ref.\cite{lan}
it is easy to show that $b_{rl} = b_l$ and $b_{lr} = b_r$.
Thus one gets
\begin{equation}
\label{b}
u(x, y, z) = b_l a 
\int\limits_{-\infty}^{\infty} 
\int\limits_{-\infty}^{\infty} 
\frac{e^{i k \sqrt{a^2 + x_1^2 + y_1^2}}}{a^2+x_1^2+y_1^2} 
\frac{e^{-i k \sqrt{z^2 + (x_1-x)^2 + (y_1-y)^2}}}
{\sqrt{z^2+(x_1-x)^2 + (y_1-y)^2}}\: dx_1 dy_1,
\end{equation}
where the additional factor $a/\sqrt{a^2+x_1^2+y_1^2}$ is the cosine of the angle
between the ray, coming from the source to the point $\{x_1, y_1, 0 \}$
and the unit vector in $z$ direction.
One can see that 
the optical lengths for all rays 
(the sum of exponents in the integrand of Eq.(\ref{b})) 
from the point source to the focus, located at
$z = a$, $x = y = 0$, are zero and
the value of the field at the focal point $u(0, 0, a) = i k$,
while the geometrical optics gives an infinite field in this point.
To find $u$ in the vicinity of the focus  one should
expand the integrand in Eq.(\ref{b}) near the point $(0, 0, a)$
assuming $x \ll a$, $y \ll a$ and $|\zeta| \ll a$, where $\zeta = z-a$. Then one gets 
\begin{equation}
\label{e1}
u(\rho, \zeta) = -\frac{k a}{2 \pi i} 
\int\limits_{0}^{2 \pi} 
\int\limits_{0}^{\infty} 
\frac{\: \exp{\left[i k \frac{\displaystyle \rho r \cos{(\varphi) - a \zeta}}{\displaystyle \sqrt{a^2+r^2}}\right]}}
{(a^2+r^2)^{3/2}}
\: r d\varphi dr,
\end{equation}
where $\rho^2 = x^2+y^2$.
Equation (\ref{e1}) gives 
\begin{equation}
\label{c}
u(\rho, \zeta) = k \left[ \frac{i \sin{(k \sqrt{\rho^2+\zeta^2})}}
{k \sqrt{\rho^2+\zeta^2}} \right. 
- \left. \int\limits_{0}^{1} J_0(k \rho \sqrt{1-s^2}) \sin{(k \zeta s)} ds \right].
\end{equation}
At $\rho = 0$ one gets an analytical expression
\begin{equation}
u(0, \zeta) =  \frac{1-\cos{(k \zeta)} + i \sin{(k \zeta)}}{ \zeta }. 
\end{equation}
Another analytical expression can be obtained at $\zeta = 0$:
$u(\rho, 0) =  i \sin( k \rho)/\rho$. The same result has been
obtained in Ref. \cite{nv}.
Figure \ref{fig2} shows dimensionless function $|u(\rho, \zeta)|^2/k^2$ 
as given by Eq.(\ref{c}).
One can see that the smearing of the focus is anisotropic.
The half-width in $z$ direction is approximately one wavelength while
in $\rho$ direction it is approximately twice as less.

Now we find the field $u$ in the close vicinity of the second focus located
at $x = y = 0$, $z = 2d-a$.
The general expression for $u$ at  $z>d$ 
differs from Eq.(\ref{b}).
One should apply the Huygens's principle  to both interfaces located at  
$z=0$ and  $z=d$.
Then the expression for the field has a form
\begin{multline}
\label{sf0}
u(x, y, z) = b_l b_r a d 
\int\limits_{-\infty}^{\infty}  \int\limits_{-\infty}^{\infty} \: dx_1 dy_1 
\int\limits_{-\infty}^{\infty}  \int\limits_{-\infty}^{\infty} \: dx_2 dy_2 
\frac{e^{i k \sqrt{a^2 + x_1^2 + y_1^2}}}{a^2+x_1^2+y_1^2}  \\
\frac{e^{-i k \sqrt{d^2 + (x_1-x_2)^2 + (y_1-y_2)^2}}}{d^2+(x_1-x_2)^2+(y_1-y_2)^2} 
\frac{e^{-i k \sqrt{(z-d)^2 + (x_2-x)^2 + (y_2-y)^2}}}{\sqrt{(z-d)^2+(x_2-x)^2+(y_2-y)^2}}.
\end{multline}
To calculate these integrals in the vicinity of the second focus
using inequalities $k d \gg 1$, $ka\gg 1$ one should introduce new variables $\{ s,t \}$
instead of variables $\{x_2, y_2\}$ by relations 
\begin{equation}
\label{lines}
x_2 = -(\frac{d}{a} - 1) x_1 + s{\rm ,}\ \ \ \ y_2 = -(\frac{d}{a} - 1) y_1 + t.
\end{equation}
One can see from Eq.(\ref{sf0}) that at $s = t = 0$ optical lengths 
of all rays exiting from the point source at $z=-a$
and coming to the second focus at $z=2d-a$ are equal to zero. 
Thus the new variables $\{ s,t \}$ describe deviation from the geometrical optics
and they should be small.
Introducing the new variables and expanding the exponents in Eq.(\ref{sf0})
one can get an expression for the field $u$ in the vicinity of the second focus.
For $\eta = z-2d+a$, $|\eta| \ll a$ one gets
\begin{equation}
\label{ff} 
u(x, y, \eta) = -u(x, y, \zeta)^*|_{\zeta=\eta},
\end{equation}
where the function $u(x, y, \zeta)$ is given by Eq.(\ref{b}).
In other words, the smearing of the $|u(x, y, \eta)|^2$ in the second focus 
is the same as the smearing in the first one.

Our result has a simple interpretation in terms
of uncertainty principle. One can see that the main
contributions to the 
integrals in Eqs.(\ref{b},\ref{sf0}) 
come from a circle of a radius $a$ in the $x$-$y$ plane around $z$ axis.
Thus, the effective aperture of the Veselago lens is of the order of $a$.
The uncertainty of the lateral momentum is of the order of $1/a$, which
gives the smearing of the foci of the order of $1/k$.

Now we compare our results with the analytical calculations 
of Pendry \cite{p2} and Ziolkowski and Heyman \cite{ziol}.
Our paper is valid for large $k$ only while in \cite{p2,ziol}
the near field region is also considered. The main difference between our
works is that in Ref.\cite{ziol} the incident spherical wave is
expanded into series of the plane waves in the $x-y$ plane. This series
contains  the evanescent waves (EW's) which are claimed to 
provide the ``superlensing''. The
matching of these plane waves at the interfaces is done independently. 
Note that 
each EW has the Poynting vector in the plane of the lens,
 which does not correspond to the problem under study. Moreover,
 the superposition of the EW's  diverges 
within three-dimensional domain $a<z<2d-a$ \cite{nv,hal,lens} at large values
 of $k$. This happen because their amplitudes increase as $\exp( k z)$ with 
positive $k$. Haldane\cite{hal} argues that the ``ultra violet cutoff'' for this problem exists, but 
it is outside macroscopic electrodynamics, so that upper value of $k$ is very large. However,  even with the cutoff the solution looks strange. The field is
exponentially  large within  the  domain  $a<z<2d-a$ which occupies all the
 space  between the two foci.
 The field in each focus  has a different limit from the left and from the
 right side. Therefore, we do  not think that such a solution means 
 a perfect focusing.

Finally, we have proposed the theory of diffraction in a system, 
consisting of the LHM and the RHM and have applied this  theory
to the calculation of the smearing of the foci of the Veselago lens.
This smearing is of the order of the wavelength.

\bibliography{lens}

\end{document}